\documentclass{camera}
\usepackage{graphicx}  

\def \hcm {\hbox {\ifmmode $ atom cm$^{-2}\else atom cm$^{-2}$\fi}}

\def \aap {A\&A}
\def \apj {ApJ}
\def \apjl {ApL}
\def \mnras {MNRAS}

\begin{document}

%
\title{Supergiant fast X-ray transients: the \emph{Swift} monitoring program}

%
\author{P.~Romano$^{1}$, L.~Sidoli$^{2}$, G.~Cusumano$^{1}$, V.~La Parola$^{1}$, J.A.~Kennea$^{3}$, S.~Vercellone$^{1}$, 
L.~Ducci$^{2}$, H.A.~Krimm$^{4}$, P.~Esposito$^{2}$,V.~Mangano$^{1}$, A.~Paizis$^{2}$, D.N.~Burrows$^{3}$ \and N.~Gehrels$^{4}$}

%
\organization{$^{1}$ INAF-IASF Palermo, Via U.\ La Malfa 153, I-90146 Palermo, Italy; \\ 
$^{2}$ INAF-IASF Milano, Via E.\ Bassini, I-20133 Milano,  Italy;  \\ 
$^{3}$ Dept.\ of Astronomy \& Astrophysics, PSU,  
PA 16802, USA; \\ 
$^{4}$ NASA/Goddard Space Flight Center, Greenbelt, MD 20771, USA 
}

\maketitle

\begin{abstract}
For the first time, {\it Swift} is giving us the opportunity to study supergiant
fast X--ray transients (SFXTs) throughout all phases of their life:  
outbursts, intermediate level, and quiescence.
We present our intense monitoring of four SFXTs,
observed 2--3 times per week since October 2007. 
We find that, unexpectedly, SFXTs spend most of their time in an 
intermediate level of accretion ($L_{X}\sim 10^{33-34} $   
erg s$^{-1}$), characterized by rich flaring activity. 
We present an overview of our investigation on SFXTs with {\it Swift},  
the key results of our Project. We  
highlight the unique contribution {\it Swift} is giving to this field, 
both in terms of outburst observations and through a systematic monitoring.  
\end{abstract}

%

\section{Supergiant fast X-ray transients} 

Supergiant fast X--ray transients (SFXTs) constitute a new class of High Mass
X--ray Binaries (HMXBs).  Discovered by {\it INTEGRAL}  \cite{Sguera2005},  
they are associated with blue supergiant stars, 
and display sporadic outbursts significantly shorter 
than those of typical Be/X-ray binaries,  characterized  
(as observed by {\it INTEGRAL}/IBIS) by bright flares lasting
a few hours with peak luminosities of 10$^{36}$--10$^{37}$~erg~s$^{-1}$ 
\cite{Sguera2005,Negueruela2006}.
The quiescence, which is characterized by a soft spectrum (likely thermal) 
and a luminosity of $\sim 10^{32}$~erg~s$^{-1}$, is a rarely-observed state
(IGR~J17544--2619, \cite{zand2005}; IGR J08408--4503, \cite{Leyder2007}). 
Therefore, SFXTs show a very large dynamic range of 3--5 orders of magnitude. 
Their hard X--ray spectra during outburst resemble the typical shape of HMXBs 
hosting accreting neutron stars (NS), with a hard power law below 10\,keV, and a high 
energy cut-off at $\sim 15$--30~keV, sometimes strongly absorbed at soft energies 
\cite{Walter2006,SidoliPM2006}.

\begin{figure}
\resizebox{\hsize}{!}{\includegraphics[height=5cm,clip=true]{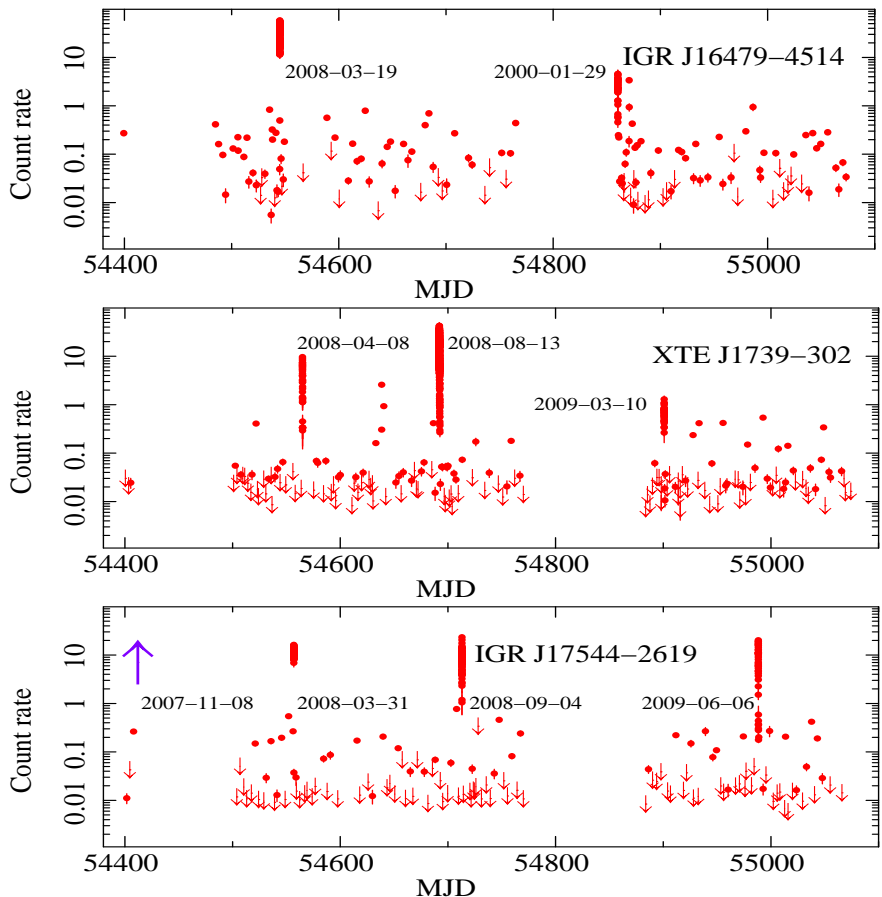}}
\vspace{-4.0truecm}
\caption{{\it Swift}/XRT (0.2--10\,keV) light curves, corrected for pile-up, 
                PSF losses, vignetting and background-subtracted. The data shown here were  
                collected from 2007 October 26 to 2009 August 31 (total of $\sim 540$ ks). 
                The reported dates refer to the BAT outbursts.
               The upward pointing arrow marks an outburst that triggered the BAT on MJD 54,414, 
                which  was not followed by the XRT due to a Sun contraint.
              		The downward-pointing arrows are 3-$\sigma$ upper limits. 
                Project web page at: { \bf http://www.ifc.inaf.it/sfxt/ .}
 }
\label{fig01} 
\end{figure}

{\it Swift}'s contribution to this field has been two-fold. 
First, it has given SFXTs the first non serendipitous 
attention throughout all phases of their life through an extensive  monitoring campaign. 
The {\it Swift} monitoring of the outburst of the periodic SFXT IGR J11215--5952 
in 2007 February \cite{Romano2007}, 
showed that the X--ray light curve is composed of several bright flares, superimposed on 
a much longer accretion phase.  
This implies that the short duration flares ($\sim$hours) observed with {\it INTEGRAL}
 and with {\it RXTE} are
actually only the brightest flares, part of a much longer outburst event lasting a few days, not only hours.
Our monitoring campaign on 4 SFXTs \cite{Sidoli2008:sfxts_paperI,Romano2009:sfxts_paperV}
since October 2007 (see Fig.~\ref{fig01}) has allowed the first assessment of how long each source 
spends in each state using a systematic monitoring with a sensitive instrument.
We discovered that X--ray emission from SFXTs is still present outside the 
bright outbursts, although at a much lower level (10$^{33}$--10$^{34}$~erg~s$^{-1}$). 
We calculated the duty-cycle of inactivity to be $\sim 17, 28, 39, 55$\,\%,  
for IGR~J16479$-$4514, AX~J1841.0$-$0536, XTE~J1739--302, and IGR~J17544$-$2619,  
respectively \cite{Romano2009:sfxts_paperV}. True quiescence is thus a rare state, 
when compared  with estimates from less sensitive instruments. 
This demonstrates that these transients accrete matter throughout their life at 
different rates. Out-of-outburst intensity-based spectroscopy shows that spectral 
fits with an absorbed blackbody result in blackbody 
radii of a few 100\,m, consistent with being emitted from a small portion 
of the NS surface, very likely the NS polar caps \cite{Romano2009:sfxts_paperV}.

Second, {\it Swift} is currently the only Observatory which can catch 
outbursts from these transients and examine them panchromatically as they 
evolve in their early
stages, thus addressing the basic issue of the nature of the mechanisms producing them. 
This has allowed us to perform the first study of truly simultaneous spectra 
(IGR~J16479--4514, \cite{Romano2008:sfxts_paperII}, in the 0.3--150 keV band) 
and fit them succesfully not only with phenomenological models 
(such as absorbed cutoff power-laws) 
but also, for the first time for SFXTs, with physical models 
such as the Comptonization model. 
The X--ray light curve of the 2008 Jul 5 outburst of IGR~J08408--4503 
\cite{Romano2009:sfxts_paper08408}
showed a multiple-peaked structure with three equally bright 
flares within 75\,ks whose spectral characteristics differ dramatically, 
with most of the difference, as derived via time-resolved spectroscopy, being due to absorbing 
column variations. 
The broad-band spectrum \cite{Sidoli2009:sfxts_paperIII} of the 2008 Sep 21 outburst 
required two distinct photon populations, a cold one ($\sim$0.3 keV) most likely 
from a thermal halo around the neutron star and a hotter one (1.4--1.8 keV) 
from the accreting column. 
We also found variable $N_{\rm H}$ in the prototype XTE~J1739--302,   
higher during an outburst than in the out-of-outburst state \cite{Sidoli2008:sfxts_paperI}, 
and varying during an outburst \cite{Sidoli2009:sfxts_paperIII}. 
The outbursts of IGR~J17544--2619 were caught both as they triggered 
the BAT \cite{Sidoli2009:sfxts_paperIII}, and as part of our monitoring campaign 
\cite{Sidoli2009:sfxts_paperIV}.

Individual sources behave somewhat differently, yet we confirmed 
in all cases the spectral similarity with accreting NS and 
highlighted the common X--ray characteristics of the class:  
{\it i)} outburst lengths well in excess of hours;
{\it ii) } outbursts with a multiple-peaked structure whose flares we used to 
confirm inhomogeneities in the wind and model the wind parameters, and 
{\it iii) } a high dynamic range (3--4 orders of magnitude in luminosity).

\noindent
{\bf Acknowledgments.} 
Grants: ASI~I/088/06/0,~I/023/05/0;~NAS5-00136.

%

\end{document}